\documentclass[aps,prl,reprint,superscriptaddress]{revtex4-2}

\usepackage{mathrsfs}
\usepackage{graphicx}
\usepackage{dcolumn} 
\usepackage{bm}
\usepackage{mathtools}
\usepackage{color}
\usepackage{braket}
\usepackage{tikz}
\usepackage{amssymb}
\usepackage{amsmath}
\usepackage{comment}

\begin{document}

\title{The Origin of Imperfection Sensitivity in the Buckling of Cylindrical Shells}

\author{Tian Yang}
\author{Tobias M. Schneider}\email{tobias.schneider@epfl.ch}

\affiliation{Emergent Complexity in Physical Systems Laboratory (ECPS), \'Ecole Polytechnique F\'ed\'erale de Lausanne, CH-1015 Lausanne, Switzerland}

% \date{\today}

\begin{abstract}
Buckling of thin cylindrical shells under axial compression, a classical example of a subcritical instability, is highly sensitive to small imperfections, with minute geometric variations causing large changes in buckling threshold. To uncover the origin of this sensitivity, we use numerical continuation and bifurcation analysis while systematically varying the depth and size of a single localized Gaussian defect. We show that the instabilities of the imperfect shell originate from localized equilibria already present in the perfect shell. By breaking translation symmetry, the defect pins these equilibria and changes how they connect to the imperfect base state. Small changes in defect geometry can thereby switch the bifurcation that triggers buckling, producing non-monotonic and discontinuous changes in buckling threshold and abrupt changes in buckling mode. Imperfection sensitivity is therefore not simply sensitivity to imperfection magnitude, but sensitivity of the underlying bifurcation structure to imperfection geometry.
\end{abstract}

\maketitle

Subcritical instabilities can render physical systems extremely sensitive to small imperfections, with minute variations producing large changes in the onset and nature of instability. Thin cylindrical shells under axial compression provide a paradigmatic example: despite their exceptional load-bearing capacity at minimal weight, nominally similar shells can buckle at vastly different loads, often far below the theoretical prediction for a perfect shell~\cite{weingarten1968buckling}. Understanding the origin of this imperfection sensitivity remains both a fundamental problem in nonlinear stability and a major challenge in structural mechanics.

Previous studies have quantified the reduction in buckling load for specific classes of imperfections~\cite{koiter1967stability,hutchinson1965axial,tennyson1969buckling, a1969effect, hutchinson1971effect, amazigo1972asymptotic, hutchinson2018imperfections, ubamanyu2026butterfly} or characterized variations among shells with random imperfections statistically~\cite{Amazigo1969, Roorda1969, ElishakoffArbocz1982, ElishakoffArbocz1985, ArboczHol1995}. Experiments further show that buckling initiates locally~\cite{Almroth1964,Tennyson1969, singer2002buckling, cuccia2023hitting}. 
Consistent with these observations, theory reveals that localized unstable equilibria in perfect shells coexist with the unbuckled state at subcritical loads and form part of its basin boundary~\cite{horak2006cylinder,kreilos2017fully,groh2019role}; their signatures can be detected by localized probing~\cite{virot2017stability, thompson2017probing, gerasimidis2018establishing, abramian2020nondestructive, cuccia2023hitting, yadav2021nondestructive, groh2023probing, lachmann2025soft}. Whether these localized equilibria provide a mechanism by which minute variations in imperfection geometry produce vastly different buckling thresholds remains unclear.

Here we show that this sensitivity can arise from competition between distinct instabilities even for a single localized imperfection characterized by only two geometric parameters. Using numerical continuation and bifurcation analysis, we systematically vary the depth and size of a Gaussian defect. Small changes in either parameter can switch the instability that triggers buckling, producing non-monotonic and discontinuous variations in the critical load and abrupt changes in buckling mode. These switches result from changes in how the imperfect base state connects to localized equilibria already present in the perfect shell, providing a deterministic mechanism for the extreme imperfection sensitivity of cylindrical shells.

We consider an elastic cylindrical shell of thickness $t$, radius $R$, and length $L$, clamped at both ends and subjected to an imposed end-shortening $\Delta$ \cite{yamaki1984elastic}. The unloaded reference mid-surface $\vec{r}_0$ contains a centered Gaussian defect of depth $\delta<0$ and size $l>0$,
\begin{equation}
    \vec r_0(x,y)  =R\vec e_r+y\vec e_a+
    \delta e^{-(x-W/2)^2/l^2-(y-L/2)^2/l^2}\vec e_r,
\label{eq:ref_config}
\end{equation}
where $x$ and $y$ are the circumferential and axial coordinates, respectively, $W = 2\pi R$, and $\vec{e}_r$ and $\vec{e}_a$ are radial and axial unit vectors. We consider physically relevant defects with $|\delta|\sim t$ and $l\sim l_c$, where $l_c$ is the classical buckling half-wavelength~\cite{timoshenko2012theory}. We compute equilibrium branches by numerical continuation of a finite-element discretization \cite{sze2004popular,betsch1998parametrization,arnold1997locking,arnold1997partial,hale2018simple} of geometrically exact shell theory~\cite{simo1989stress,simo1990stress,bischoff2004models} and determine their stability and bifurcations from the spectrum of the Jacobian \cite{dalcinpazklercosimo2011,hernandez2005slepc,baratta2023dolfinx,fenicsx-shells_2025} (see Supplemental Material~\cite{supplemental_material}).

\begin{figure*}[htbp!]
    \centering
    \includegraphics[width=1\linewidth]{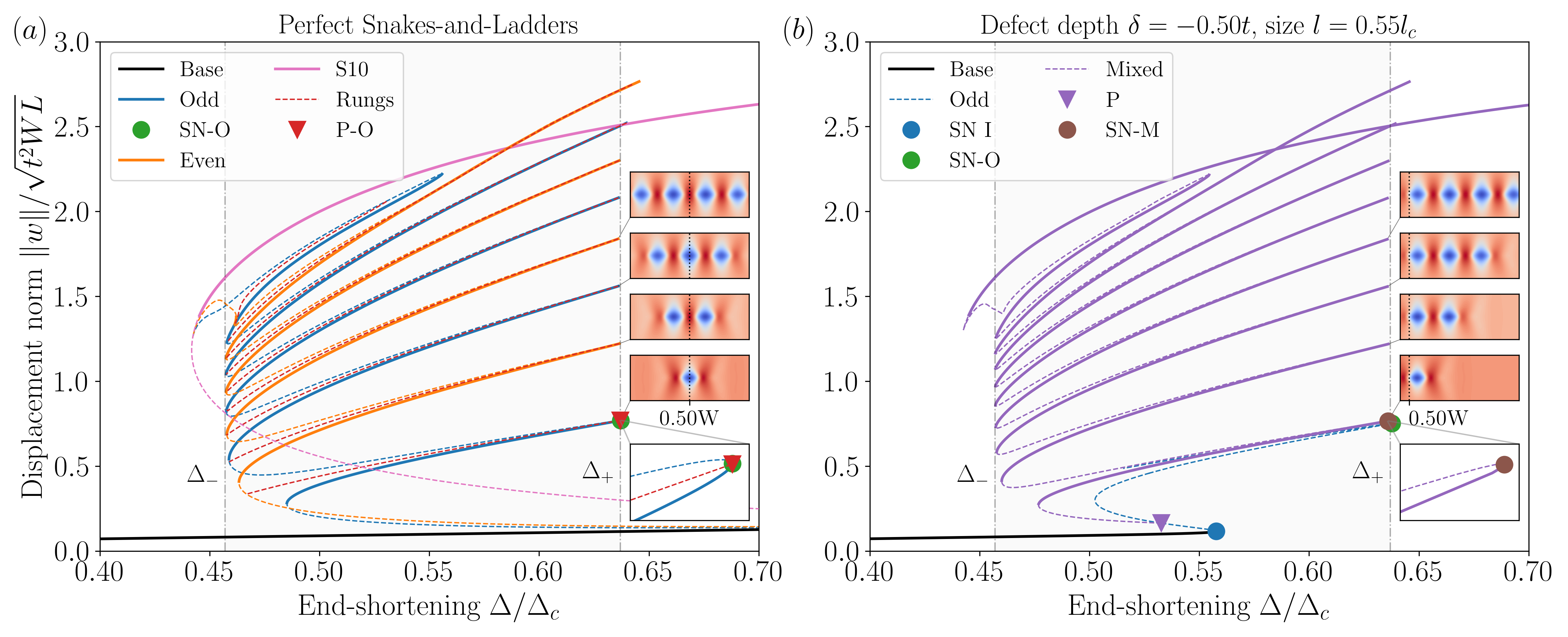}
    \caption{
    Bifurcation diagrams of the normalized radial-displacement norm, $\|w\|/\sqrt{t^2WL}$, versus normalized end-shortening, $\Delta/\Delta_c$.
    (a) Perfect shell: base state, odd and even snaking branches, spatially periodic ten-dimple branch (S10), and rung branches. Green circles and red triangles mark saddle-node (SN-O) and pitchfork (P-O) bifurcations on the odd branch.
    (b) Imperfect shell with $\delta=-0.50t$ and $l=0.55l_c$: base state, odd snaking branch, and mixed snaking branches. Blue circles, purple triangles, green circles, and brown circles mark SN~I, P, SN-O, and SN-M, respectively.
    Thick solid and thin dashed curves denote stable and unstable equilibria. Insets show representative radial-displacement fields at $\Delta_+$; lower insets enlarge the bifurcation diagrams near $\Delta_+$.
}
    \label{fig:1}
\end{figure*}

In the subcritical regime of the perfect shell ($\delta=0$), the stable prebuckling state coexists with localized equilibria consisting of dimples along the shell's midsection. These states form the characteristic snakes-and-ladders structure~\cite{kreilos2017fully,groh2019role} shown in Fig.~\ref{fig:1}(a). Two branches containing odd and even numbers of dimples snake within $(\Delta_-,\Delta_+)\approx(0.457\Delta_c,0.637\Delta_c)$,
where $\Delta_c$ is the classical critical end-shortening \cite{supplemental_material}. Successive saddle-node bifurcations at $\Delta_+$ nucleate pairs of dimples at the two fronts of the localized pattern until both branches terminate on the spatially periodic S10 branch. States on the odd and even branches are reflection symmetric about the center of one dimple and the midpoint between two dimples, respectively. Reflection-symmetry-breaking pitchfork bifurcations near $\Delta_+$ generate asymmetric rung branches that nucleate a dimple at only one front and connect the odd and even snaking branches~\cite{groh2021snaking}.

The perfect shell has the same relevant spatial symmetries as the Swift--Hohenberg equation with quadratic-cubic nonlinearity~\cite{burke2006localized,burke2007homoclinic,knobloch2015spatial}: continuous azimuthal translation symmetry and reflection symmetry. Translation symmetry implies that every localized state belongs to a continuous one-parameter family of translated copies, with an associated neutral shift mode. It also permits asymmetric rungs to connect the odd and even snaking branches, whose reflection symmetries are defined about different origins.
A localized defect breaks the translation symmetry, lifting this degeneracy and selecting isolated, defect-pinned equilibria from the continuous families of translated states~\cite{kao2014spatial}. In the following, we focus on the pinned branches connected to the prebuckling state and hence directly relevant to buckling.

For a representative defect with $\delta=-0.50t$ and $l=0.55l_c$, the prebuckling state contains a single reflection-symmetric dimple pinned at the defect [Fig.~\ref{fig:1}(b)]. It loses stability at the saddle-node bifurcation SN~I at $\Delta_{\rm SNI}\approx0.558\Delta_c$ and connects to the reflection-symmetric odd snaking branch. The snaking interval is only weakly affected because the saddle-node eigenmodes are localized at the fronts of the pattern, away from the defect.
The odd branch loses reflection symmetry in the pitchfork bifurcation P at $\Delta_\mathrm{P}\approx0.533\Delta_c$. Its critical eigenmode shifts the localized pattern laterally relative to the defect, generating a pair of reflection-related asymmetric branches. Along either branch, one outer dimple remains pinned while the pattern grows at the opposite front, producing a mixed snaking branch containing both odd and even numbers of dimples.

This bifurcation structure follows directly from the symmetry breaking introduced by the defect. 
Reflection symmetry about the defect is preserved for the imperfect shell. When the central dimple of the odd branch is centered on the defect, the odd branch can retain this symmetry, whereas the even branch cannot because its reflection center lies between dimples.
The rung-generating pitchforks of the perfect shell are therefore unfolded into saddle nodes connecting remnants of the rungs to the former even and odd snaking branches, producing the mixed branches. The perfect-shell snakes-and-ladders structure is recovered continuously as $\delta\to0$.

The defect primarily affects the formerly neutral shift mode. On the reflection-symmetric odd branch this mode becomes unstable at P, whereas on the mixed branch it is stable; the remaining stability properties are essentially inherited from the perfect-shell branches, with stability changes occurring at their saddle nodes.

\begin{figure*}[htbp]
    \includegraphics[width=1\linewidth]{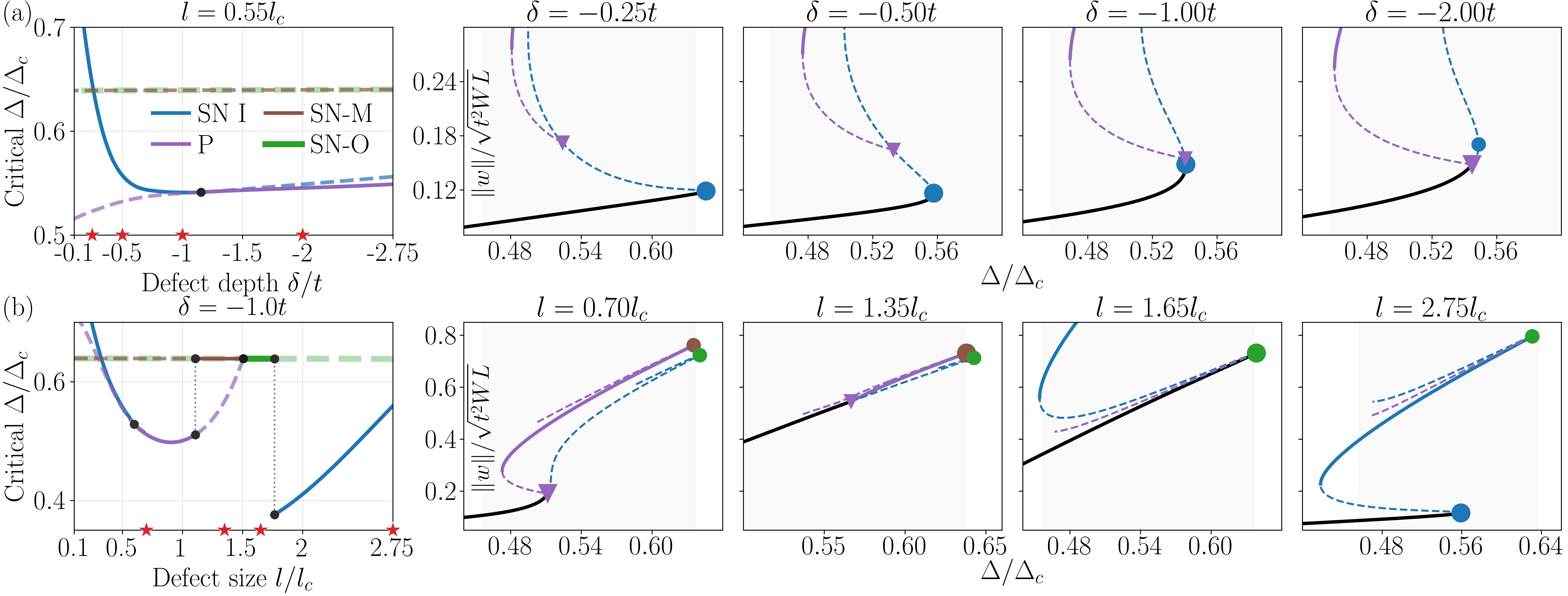}
    \caption{
    (a) Normalized critical end-shortening $\Delta/\Delta_c$ versus defect depth $\delta/t$ at fixed $l/l_c=0.55$, with representative bifurcation diagrams for $\delta/t=-0.25,-0.50,-1.00$, and $-2.00$.
    (b) Corresponding dependence on defect size $l/l_c$ at fixed $\delta/t=-1.0$, with bifurcation diagrams for $l/l_c=0.70,1.35,1.65$, and $2.75$.
    Blue, purple, brown, and green curves denote SN~I, P, SN-M, and SN-O, respectively. Solid curves identify the bifurcation that triggers buckling; dashed curves show the remaining bifurcations. Black markers indicate transitions between primary buckling bifurcations, with dotted lines in (b) marking discontinuous changes in the buckling threshold. Red stars mark the parameter values of the representative bifurcation diagrams, whose line and marker styles follow Fig.~\ref{fig:1}; the largest marker identifies the primary instability.
}
    \label{fig:2}
\end{figure*}

In summary, the imperfect shell possesses two types of stable equilibria: the reflection-symmetric base state and localized states on the mixed snaking branches. For the representative defect, the base state loses stability at SN~I at $\Delta_{\rm SNI}\approx0.558\Delta_c$, whereas P occurs on the subcritical odd branch at $\Delta_\mathrm{P}\approx0.533\Delta_c$. The mixed branches become stable only at a subsequent saddle node, yielding a finite interval of coexistence with the stable base state, and remain stable beyond SN~I up to SN-M at $\Delta_+\approx0.636\Delta_c$.

The bifurcation structure is directly reflected in the mechanical response: at SN~I, loss of stability of the prebuckling state is accompanied by an abrupt drop in reaction force (Fig.~S5(b) of~\cite{supplemental_material}). Although stable localized postbuckling states coexist, their dynamical selection is not determined by the static bifurcation diagram alone. We therefore identify the first loss of stability of the prebuckling state under increasing end-shortening as the buckling threshold.

To determine how the bifurcation structure and the associated buckling behavior depend on the imperfection parameters, we first fix the defect size, $l=0.55l_c$, and vary its depth $\delta$ [Fig.~\ref{fig:2}(a)]. As $\delta\rightarrow0$, the critical end-shortening of the symmetry-preserving SN~I approaches that of the perfect shell. Increasing the defect depth initially reduces this threshold, which approximately saturates once $|\delta|/t=O(1)$ and increases slightly for still deeper imperfections. In contrast, the critical end-shortening of the symmetry-breaking P increases monotonically with $|\delta|$. At $\delta=-1.16t$, P and SN~I meet in a codimension-two bifurcation. Beyond this point, P moves from the unstable to the stable side of SN~I and therefore becomes the primary instability of the base state. Increasing the defect depth thus changes both the critical end-shortening and the nature of the primary buckling bifurcation, from a symmetry-preserving saddle node to a symmetry-breaking pitchfork.

The dependence on defect size is considerably richer. For fixed depth $\delta=-1.0t$ and small $l$, SN~I remains the primary instability, with P located on its subcritical unstable branch [Fig.~\ref{fig:2}(b)]. At $l=0.60l_c$, the two bifurcations meet at a codimension-two point and P moves onto the stable base-state branch, becoming the primary instability. Upon further increasing $l$, the pair of saddle-node bifurcations on the reflection-symmetric base-state branch annihilates at $l=0.69l_c$, leaving P as the only instability of this branch. At $l=1.11l_c$, P changes from subcritical to supercritical. Crossing it then produces a smooth transition onto a stable mixed branch rather than buckling, and the first loss of stability instead occurs at the subsequent saddle node on the mixed branch, SN-M. With increasing $l$, the pitchfork shifts to larger $\Delta$ until it collides at $l=1.51l_c$ with SN-O on the reflection-symmetric odd branch. Beyond this point, symmetry breaking no longer precedes buckling and the shell instead loses stability directly through SN-O on the reflection-symmetric odd branch. At still larger defect sizes, $l>1.77l_c$, a subcritical symmetry-preserving SN~I re-emerges as the primary instability.

\begin{figure}[htbp]
    \includegraphics[width=1\linewidth]{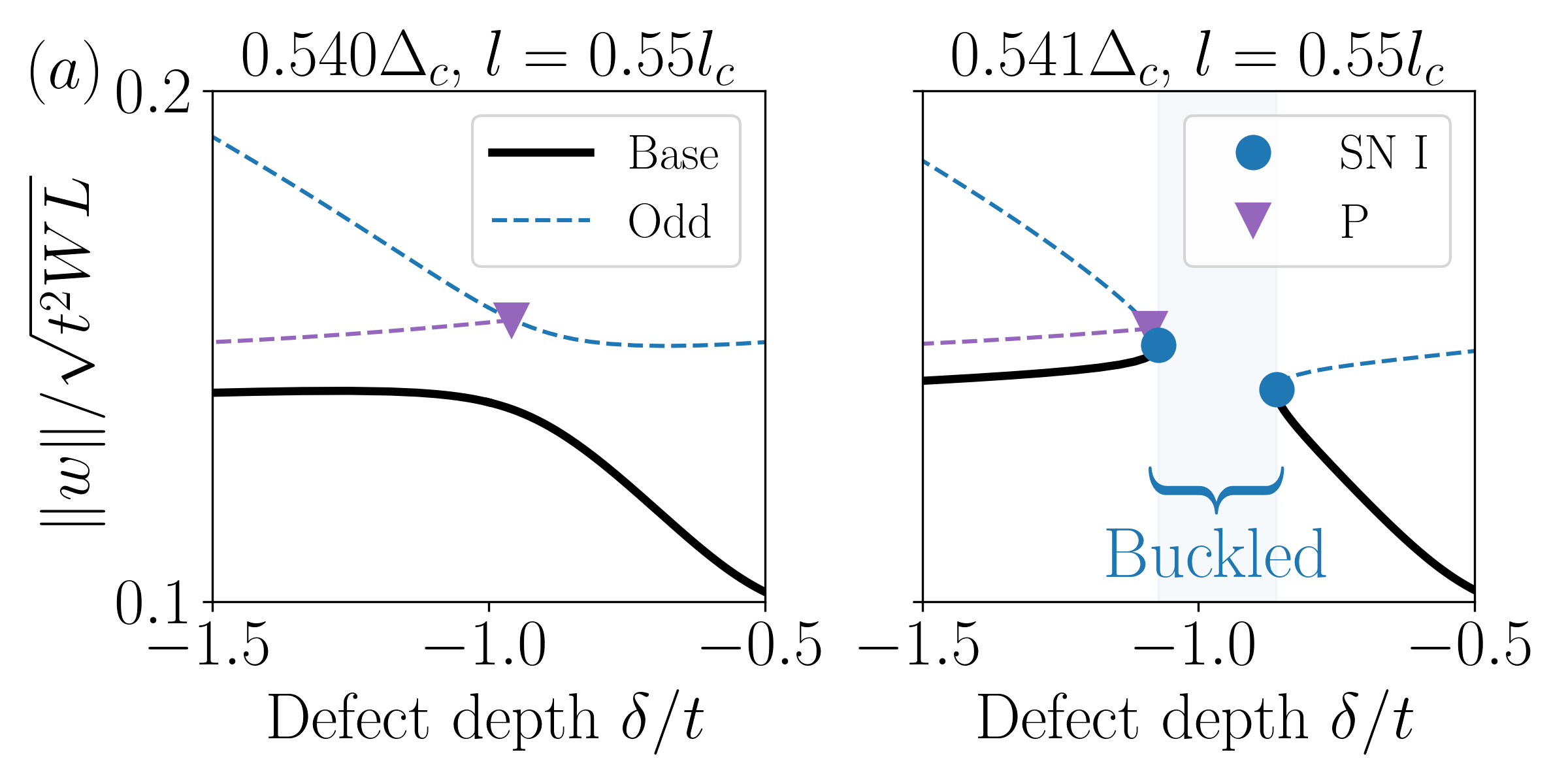}
    \includegraphics[width=0.98\linewidth]{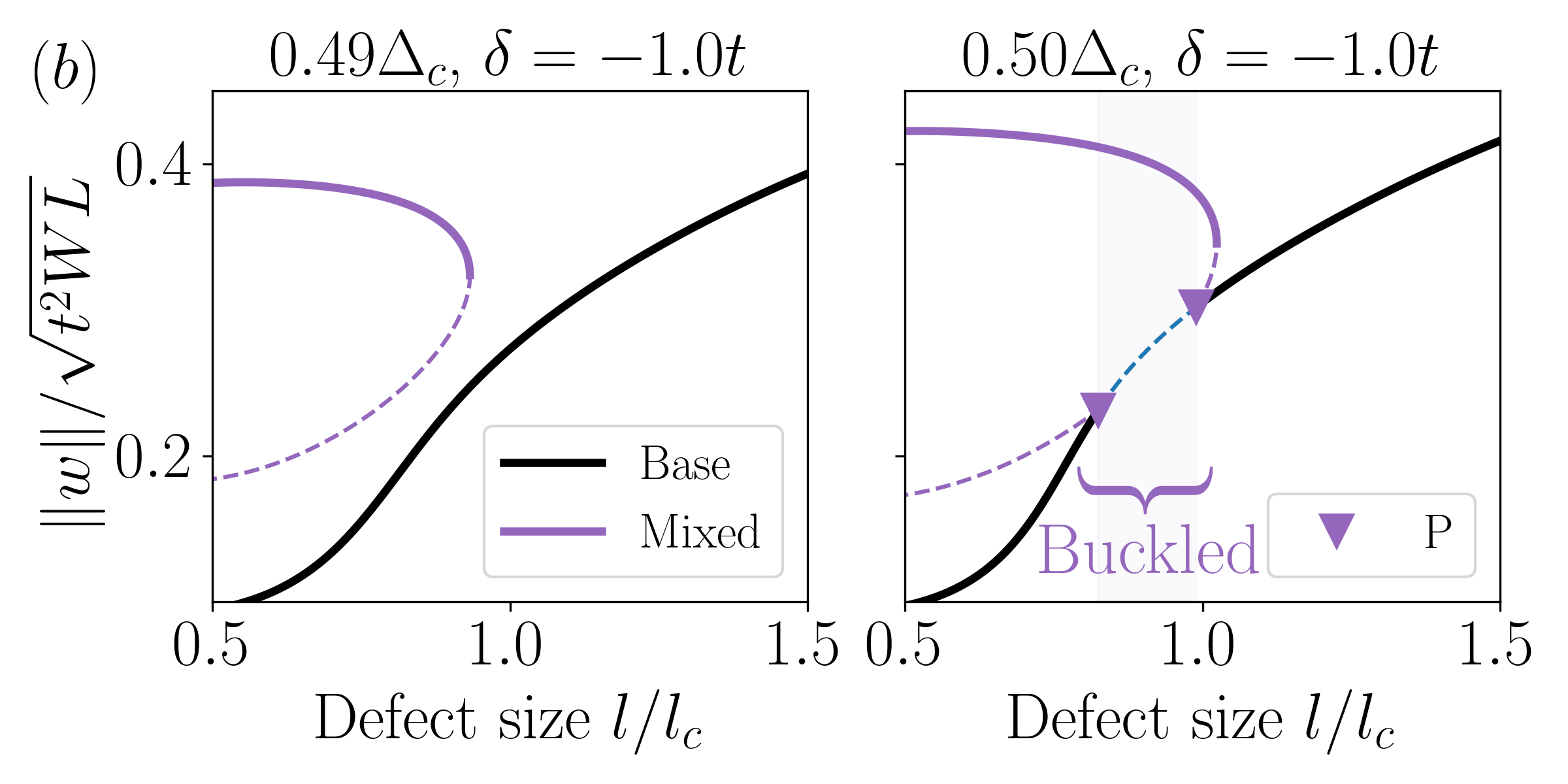}
    \caption{Bifurcation diagrams obtained by continuation in (a) defect depth $\delta/t$ at fixed $l/l_c=0.55$ for $\Delta/\Delta_c=0.540$ and $0.541$, and (b) defect size $l/l_c$ at fixed $\delta/t=-1.0$ for $\Delta/\Delta_c=0.49$ and $0.50$. The SN~I bifurcations in (a) bound an interval where the base branch is absent, whereas the P bifurcations in (b) bound an interval where it is unstable. These intervals widen as $\Delta$ increases, confirming the non-monotonic dependence of $\Delta_{\mathrm{SNI}}$ on $\delta$ and of $\Delta_{\mathrm{P}}$ on $l$.}
    \label{fig:3}
\end{figure}

The critical end-shortening therefore depends non-monotonically on defect size. While SN I and P vary strongly with $l$, SN-M and SN-O remain near the upper snaking limit $\Delta_+$ inherited from the perfect shell. Exchanges between these primary bifurcations produce non-smooth variations in both the buckling threshold and mode.
Direct continuation in the imperfection parameters confirms the non-monotonic dependence of the buckling threshold. For fixed $l=0.55l_c$, the stable base state exists for all $\delta$ at $\Delta=0.540\Delta_c$, whereas at the slightly larger end-shortening $\Delta=0.541\Delta_c$ it is absent over a finite interval around $\delta/t\approx-1$; shells with both shallower and deeper imperfections remain stable [see Fig.~\ref{fig:3}(a)]. Similarly, for fixed $\delta=-1.0t$ at $\Delta=0.50\Delta_c$, buckling occurs only for $l/l_c\in[0.82,0.99]$, while shells with both smaller and larger defect sizes remain stable [see Fig.~\ref{fig:3}(b)].

%\section{Buckling over two defect parameters}
\begin{figure}[ht!]
    \centering
    \includegraphics[width=1\linewidth]{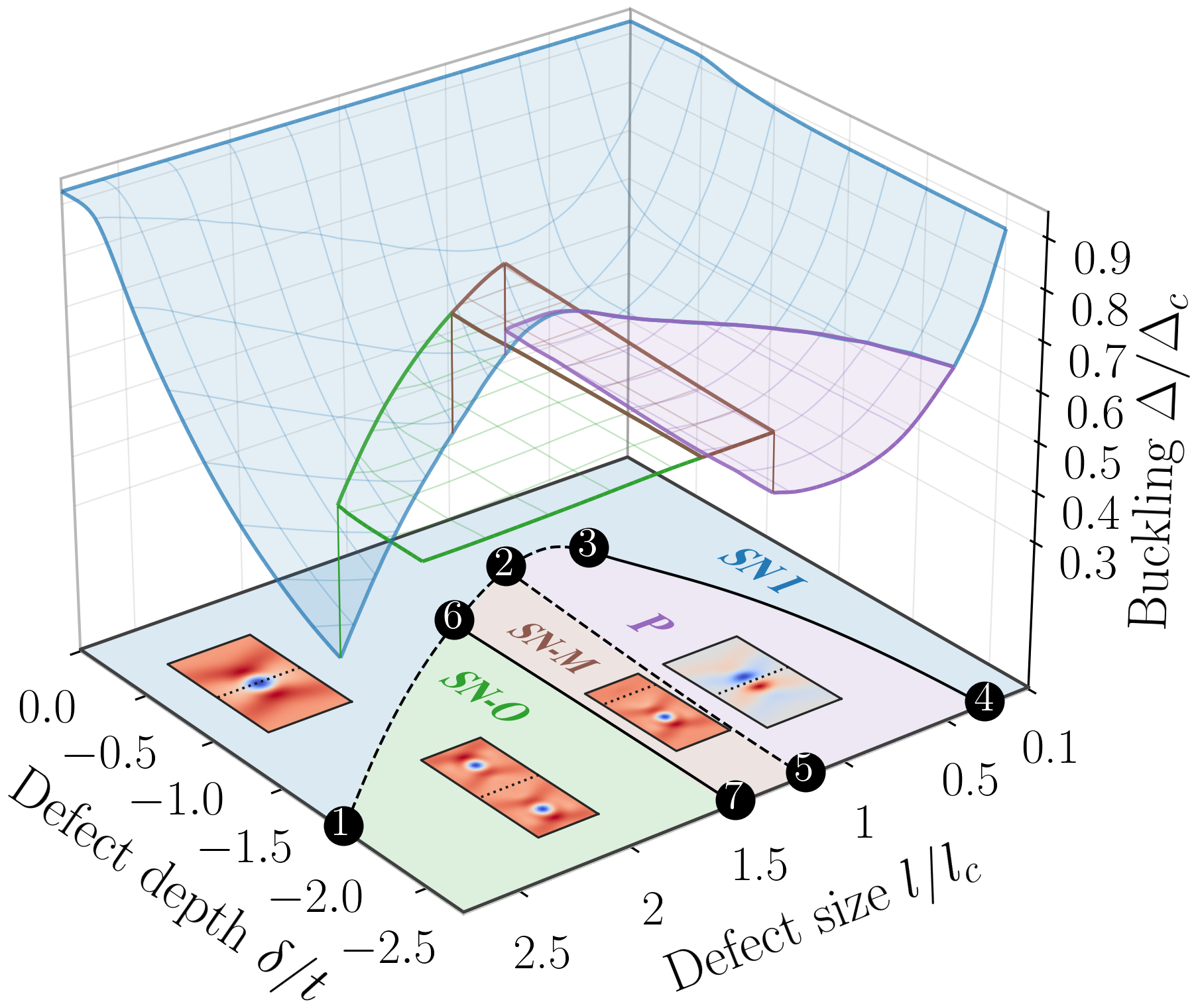}
    \caption{The colored defect-parameter plane $(\delta/t,l/l_c)$ identifies the primary buckling bifurcation in each region; the insets show representative eigenmodes with the dotted line indicating the defect location. The normalized buckling end-shortening, $\Delta/\Delta_c$, is plotted as color-matched surfaces. Markers 1--7 divide the region boundaries into six segments. Segments 1--6, 6--2 and 2--3 indicate the existence of SN~I, while segment 3--4 indicates the relative position of SN~I and P. Segment 2--5 marks where P changes between subcritical and supercritical; when P is supercritical, buckling instead occurs at SN-M. Segment 6--7 marks the disappearance of P, beyond which buckling occurs at SN-O on the reflection-symmetric odd branch.}
    \label{fig:4}
\end{figure}

The sensitivity to the detailed defect geometry becomes apparent when considering the full two-dimensional parameter space $(\delta,l)$. Figure~\ref{fig:4} shows the critical end-shortening together with the associated buckling bifurcation, revealing four regions in which the primary instability occurs at SN~I, P, SN-M, or SN-O. Within the SN~I and P regions, the critical end-shortening varies strongly and non-monotonically with both defect parameters, whereas buckling at SN-M and SN-O occurs close to the upper snaking limit $\Delta_+$. Depending on how the underlying bifurcations meet, transitions between these regions are either continuous or produce abrupt jumps in the buckling threshold. This complex dependence occurs already for defect sizes $l/l_c=O(1)$ and depths $|\delta|/t=O(1)$.

Taken together, our results show that large variations in cylindrical-shell buckling strength do not require complex or randomly distributed imperfections or interactions among multiple defects. Even a single localized defect characterized by only two geometric parameters produces non-monotonic and discontinuous buckling thresholds, as small changes in its geometry switch the bifurcation that triggers buckling. Imperfection sensitivity is therefore not simply sensitivity of the buckling threshold to imperfection magnitude, but sensitivity of the bifurcation structure to imperfection geometry. This decouples geometric magnitude from mechanical severity: deeper or larger defects are not necessarily more detrimental, and even within this minimal two-parameter family no obvious scalar measure of imperfection magnitude ranks them by their impact on buckling. Moreover, symmetry-breaking primary instabilities imply that equilibrium-path calculations that identify buckling solely with limit points, or analyses restricted to symmetry-preserving deformations, can miss the relevant buckling threshold. Thus, the extreme sensitivity is intrinsic to the nonlinear shell mechanics: local variations in defect geometry are amplified through changes in the global bifurcation structure into large variations in buckling strength.

\bibliography{reference}

\end{document}

%Title of paper
\title{Supplemental Materials}
\maketitle

\setcounter{section}{0}
\renewcommand{\thesection}{S\arabic{section}}
\renewcommand{\theequation}{S\arabic{equation}}
\renewcommand{\thefigure}{S\arabic{figure}}
\renewcommand{\thetable}{S\arabic{table}}

\section{\centerline{1. Problem setup}}
\begin{figure}[ht!]
    \centering
    \includegraphics[width=0.99\linewidth]{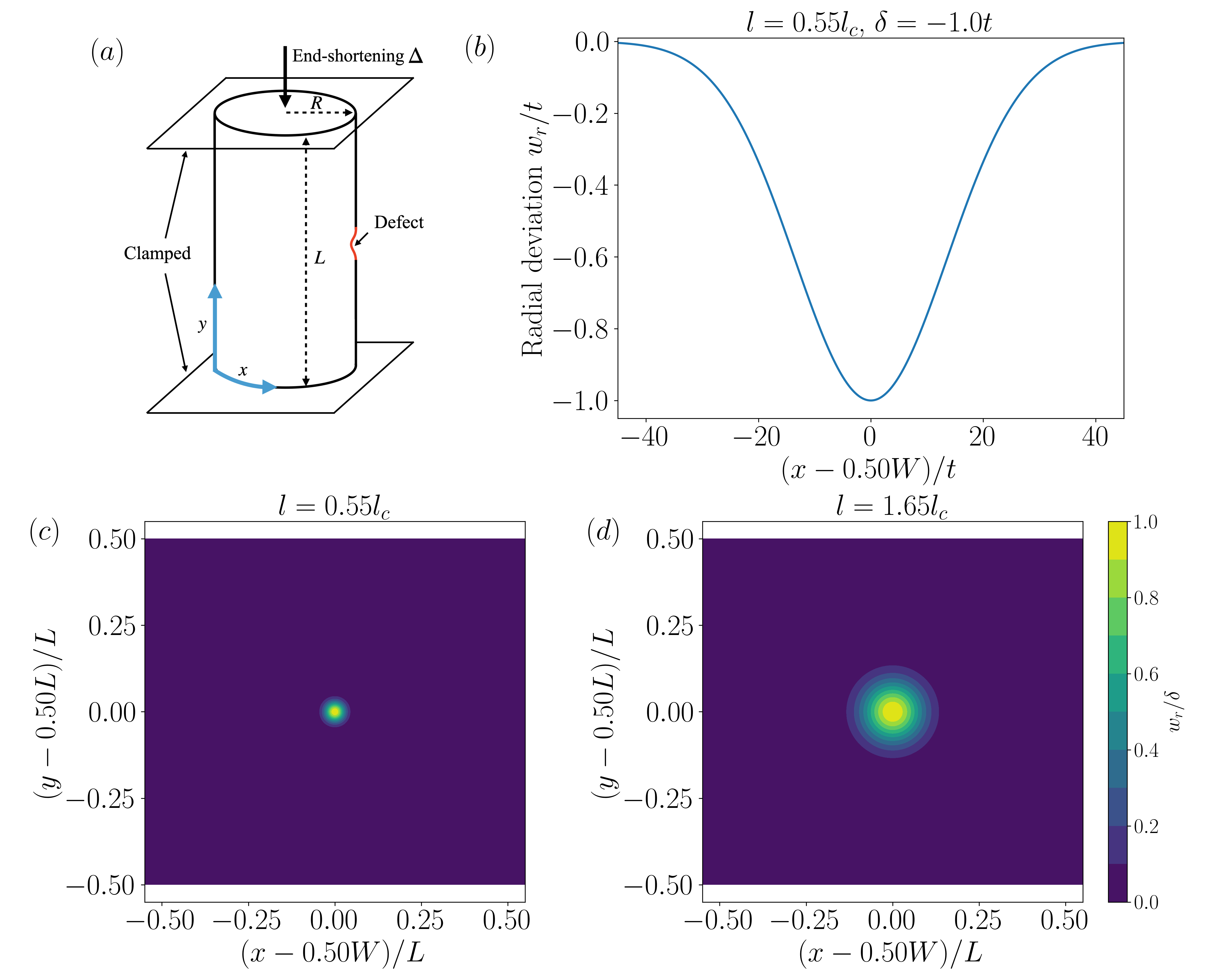}
    \caption{(a) Geometry and boundary conditions of a cylindrical shell of radius \(R\) and length \(L\), clamped at both ends and compressed by a prescribed end-shortening \(\Delta\). The circumferential and axial coordinates are denoted by \(x\) and \(y\), respectively. (b) Circumferential profile through the center of the Gaussian defect for \(l=0.55l_c\) and \(\delta=-1.0t\). Here, \(w_r\) denotes the radial deviation of the undeformed imperfect mid-surface from the perfect shell, and \(W=2\pi R\) is the circumferential width of the unrolled shell. (c),(d) Spatial distributions of the normalized radial deviation \(w_r/\delta\) around the defect center \((x,y)=(0.50W,0.50L)\) for \(l=0.55l_c\) and \(1.65l_c\), respectively. In (b), the circumferential coordinate is normalized by the shell thickness \(t\). In (c),(d), both coordinates are normalized by \(L\) to maintain a common spatial scale.}
    \label{fig:S1}
\end{figure}

The geometric and material parameters of the cylindrical shell shown in Fig.~\ref{fig:S1}(a) are listed in Tab.~\ref{tab:S1}. These parameters originate from the seminal experiments of Yamaki \cite{yamaki1984elastic}, and have been widely employed in previous studies of the snakes-and-ladders bifurcation structure of perfect cylindrical shells \cite{kreilos2017fully,groh2019role, groh2021snaking, groh2023probing}.

Linear stability analysis of a perfect cylindrical shell with simply supported boundaries yields the classical critical end-shortening $\Delta_c$, critical buckling load $P_c$, and  half-wavelength $l_c$ of the axisymmetric buckling mode \cite{koiter1967stability,timoshenko2012theory}:
\begin{equation}
    \Delta_c=\frac{tL}{R\sqrt{3(1-\nu^2)}}, \quad P_c=\frac{2\pi Et^2}{\sqrt{3(1-\nu^2)}}, \quad  l_c = \pi \sqrt{\frac{Rt}{\sqrt{12(1-\nu^2)}}}.
\end{equation}
These classical values are used throughout as normalization scales; their numerical values are listed in Tab.~\ref{tab:S1}. For the clamped boundaries considered here, the perfect-shell buckling threshold is slightly lower, $\Delta \approx 0.93\Delta_c$ and $P \approx 0.93P_c$~\cite{groh2019role}.

\begin{table}[h!]
    \centering
    \begin{tabular}{|c|c|c|c|} \hline 
         Thickness $t$&  Radius $R$& Length $L$ & Perfect critical end-shortening $\Delta_c$\\ \hline 
         $0.247mm$&  $100mm$& $160.9mm$& $0.241mm$\\ \hline 
         Young's Modulus $E$&  Poisson ratio $\nu$& Classical half-wavelength $l_c$& Perfect critical load $P_c$\\ \hline 
         $5.56GPa$&  $0.3$& $8.59mm$& $1290N$\\ \hline
    \end{tabular}
    \caption{Geometry and material parameters of the cylindrical shell.}
    \label{tab:S1}
\end{table}

\section{\centerline{2.  Equations and numerical implementation}}

We describe the shell mid-surface using the Lagrangian circumferential and axial coordinates $(x,y)\in[0,W]\times[0,L]$, where $W=2\pi R$. The position of the mid-surface in the deformed configuration is given by the Eulerian coordinates $\vec r(x,y)\in\mathbb{R}^3$. Relative to the unloaded reference configuration $\vec r_0(x,y)$ defined in Eq.~(1) of the main text, the displacement field is
\begin{equation}
    \vec u(x,y)=\vec r(x,y)-\vec r_0(x,y).
\end{equation}
The radial component of the displacement is denoted by $w=\vec u\cdot\vec e_r$.

The nonlinear deformation of the shell is modeled using a geometrically exact shell theory that accounts for finite displacements and rotations  \cite{simo1989stress, simo1990stress}. In this theory, a general shell is represented by a Cosserat surface equipped with a single inextensible director, therefore the current Eulerian position $\vec{q}\in\mathbb{R}^3$ of a material point is given by
\begin{equation}
\label{eq:current_conf}
\vec{q}(\xi^1, \xi^2, \xi^3) = \vec{r}(\xi^1, \xi^2)+ \xi^3\vec{d}(\xi^1,\xi^2), \quad \xi^3 \in \Big[\tfrac{-t}{2}, \tfrac{t}{2}\Big],
\end{equation}
where $(\xi^1, \xi^2)$ are the Lagrangian curvilinear coordinates of the mid-surface $\vec{r}$, and $\xi^3$ is the coordinate along the unit director $\vec{d}$. For the cylindrical shells, we have the coordinates $x = \xi^1$ and $y=\xi^2$.

Based on this kinematic assumption, the metric tensor $\boldsymbol{a}$ and the curvature tensor $\boldsymbol{b}$ of the mid-surface are expressed in terms of the covariant basis vectors $\vec{a}_\alpha = \partial_\alpha \vec{r}=\partial \vec{r}/{\partial \xi^\alpha}, (\alpha = 1,2) $:
\begin{equation}
    \label{eq:first_second_fund_forms_current}
    \begin{aligned}
        a_{\alpha\beta} &=  \vec{a}_\alpha \cdot \vec{a}_\beta, \\
        b_{\alpha\beta} &= \frac{1}{2}\left ( 
    \partial_\alpha \vec{d} \cdot \vec{a}_\beta +
    \partial_\beta \vec{d} \cdot \vec{a}_\alpha
    \right ).
    \end{aligned}
\end{equation}
For the undeformed reference configuration, we use the symbols $\vec{r}_0, \vec{n}, \vec{A}_\alpha, A_{\alpha\beta}, B_{\alpha\beta}$ to replace $\vec{r}, \vec{d}, \vec{a}_\alpha, a_{\alpha\beta}, b_{\alpha\beta}$.

The Green-Lagrange strain tensor $\boldsymbol{E}$ is defined using the 3D metric tensors of the current and initial configurations:
\begin{equation}
    \label{eq:green_lagrange_strain}
        E_{ij} = \frac{1}{2}\left(g_{ij} - G_{ij}\right), \quad
        g_{ij} = \vec{g}_i \cdot \vec{g}_j, \quad 
        G_{ij} = \vec{G}_i \cdot \vec{G}_j,
\end{equation}
where $\vec{g}_i = \partial \vec{q}/\partial \xi^i$ $(i=1,2,3)$ denotes covariant basis vectors in the current configuration, while $\vec{G}_i$ represents the counterparts in the reference configuration. Following the kinematic assumption (Eq.~\ref{eq:current_conf}), these basis vectors satisfy:
\begin{equation}
\label{eq:covariant_basis_3D}
    \begin{aligned}
    \vec{G}_\alpha &= \vec{A}_\alpha + \xi^3 \partial_\alpha \vec{n}, \quad \vec{G}_3 = \vec{n}, \\
    \vec{g}_\alpha &= \vec{a}_\alpha + \xi^3 \partial_\alpha \vec{d}, \quad \vec{g}_3 = \vec{d}.
    \end{aligned}
\end{equation}
Taking Eq.~\ref{eq:covariant_basis_3D} into Eq.~\ref{eq:green_lagrange_strain} and neglecting terms that are quadratic in $\xi^3$, the components of the 3D Green-Lagrange strain tensor can be expressed solely in terms of the in-plane coordinates $(\xi^1,\xi^2)$:
\begin{equation}
    \label{eq:reduced_GL_strain}
        E_{\alpha\beta}  =  \varepsilon_{\alpha\beta}  + \xi^3 \kappa_{\alpha\beta}, \quad
        E_{\alpha 3}  = \frac{1}{2}\gamma_\alpha , \quad
        E_{33}  = 0.
\end{equation}
Here, the measures of membrane strain $\boldsymbol{\varepsilon}$ , bending strain $\boldsymbol{\kappa}$, and transverse shear strain $\vec{\gamma} $ are given by:
\begin{equation}
    \label{eq:shell_strains}
    \begin{aligned}
        \varepsilon_{\alpha\beta} &= \frac{1}{2}\left( a_{\alpha\beta} - A_{\alpha\beta} \right), \\
    \kappa_{\alpha\beta} &= b_{\alpha\beta} - B_{\alpha\beta},
    \\
    \gamma_{\alpha} &= \vec{a}_\alpha \cdot \vec{d} - \vec{A}_\alpha \cdot \vec{n}.
    \end{aligned}
\end{equation}
Consequently, the elastic strain energy contains membrane, bending, and transverse-shear contributions. Equilibrium states are stationary points of the total potential energy:
\begin{equation}
\delta\Pi_\text{total} = 0; \quad
   \Pi_{\text{total}}  =
    \frac{1}{2}\int_\Omega \left(\boldsymbol{N}:\boldsymbol{\varepsilon}  
    +  \boldsymbol{M}:\boldsymbol{\kappa} 
    +  \vec{T}\cdot\vec{\gamma}\right) \mathrm{~d}\Omega - \Pi_\text{external},
    \label{eq:potential_energy}
\end{equation}
where $\boldsymbol{N}, \boldsymbol{M}, \vec{T}$ are the energy-conjugated stress resultants. Their explicit expressions and the associated constitutive relations can be found in \cite{bischoff2004models,hale2018simple}.

The director $\vec{d}$ is parameterized by two rotational degrees of freedom, excluding drilling rotations \cite{betsch1998parametrization}. The weak form in Eq.~\ref{eq:potential_energy} is discretized using quadratic Lagrange elements. Membrane and shear locking are alleviated by partial selective reduced integration \citep{arnold1997locking,arnold1997partial}. Periodic boundary conditions are imposed in the circumferential direction. The top and bottom edges are clamped, and the end-shortening $\Delta$ is imposed through a prescribed displacement of the top edge. 

Equilibrium states are obtained by Newton iterations using the parallel numerical library PETSc~\cite{dalcinpazklercosimo2011}. The eigenspectrum of the Jacobian matrix is computed using SLEPc~\cite{hernandez2005slepc} to determine stability and identify bifurcations. Equilibrium branches are tracked as the end-shortening $\Delta$, defect depth $\delta$, or defect size $l$ is varied using a predictor--corrector continuation method. At each continuation step, the generalized displacement field and control parameter are predicted by quadratic extrapolation from preceding solutions. The control parameter is then held fixed while the displacement field is corrected by Newton iterations. The computational framework has been validated against the benchmark problems in Ref.~\cite{sze2004popular}. The shell model is implemented using the open-source FEniCS finite-element framework~\cite{hale2018simple,baratta2023dolfinx}. A documented implementation is available in the FEniCSx-shells repository~\cite{fenicsx-shells_2025}.

\section{\centerline{3. Eigenmodes and equilibrium states}}

\begin{figure*}[ht!]
    \includegraphics[width=0.98\linewidth]{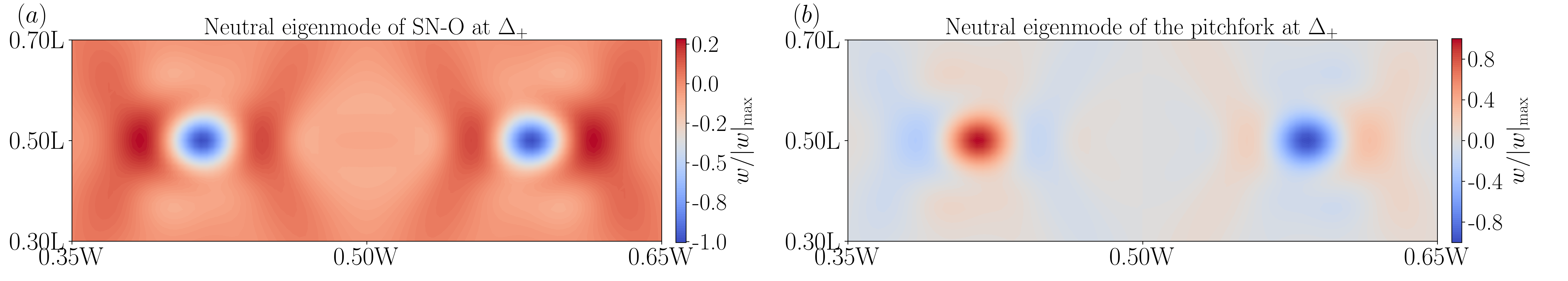}
    \caption{Neutral eigenmodes associated with (a) SN-O and (b) the adjacent P-O bifurcation at the first upper turning point of the perfect odd snaking branch, $\Delta=\Delta_+$.}
    \label{fig:S2}
\end{figure*}

Figures~\ref{fig:S2}(a) and~\ref{fig:S2}(b) show the neutral eigenmodes associated with SN-O and the adjacent P-O bifurcation on the perfect odd snaking branch in Fig.~1(a) of the main text, respectively.
The neutral eigenmode of SN-O is symmetric about the midpoint of the central dimple and nucleates a pair of dimples at both fronts. By contrast, the neutral eigenmode of P-O is antisymmetric about the central dimple and nucleates only one dimple at one front. The neutral eigenmodes of SN-O and P-O remain nearly unchanged on imperfect reflection-symmetric odd branches because they are only weakly affected by the central defect.

\begin{figure*}[ht!]
    \includegraphics[width=\linewidth]{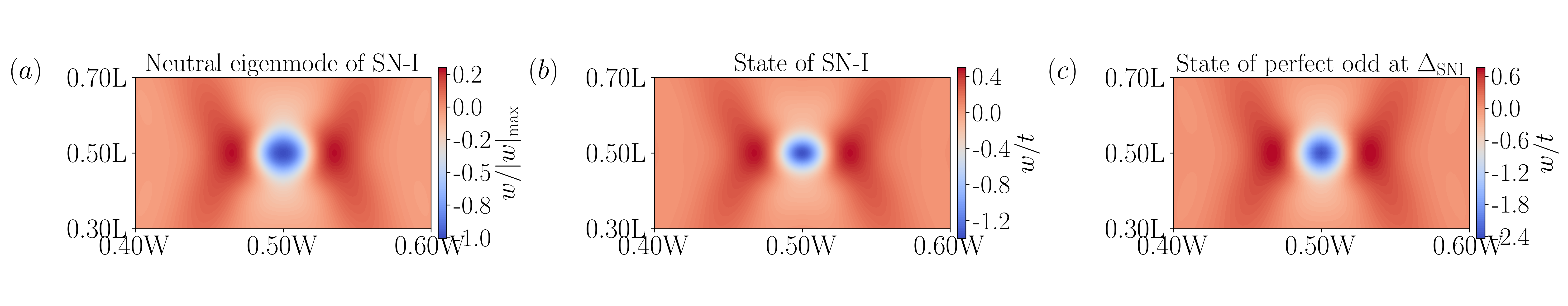}
    \caption{(a) Neutral eigenmode and (b) equilibrium state at SN~I for an imperfect shell with $\delta=-0.50t$ and $l=0.55l_c$. (c) Unstable single-dimple equilibrium on the perfect odd snaking branch at the same end-shortening, $\Delta_{\mathrm{SNI}}\approx0.558\Delta_c$.}
    \label{fig:S3}
\end{figure*}

Figures~\ref{fig:S3}(a) and~\ref{fig:S3}(b) show the neutral eigenmode and equilibrium state at SN~I for the imperfect shell of $\delta=-0.50t$, $l=0.55l_c$, respectively. For comparison, Fig.~\ref{fig:S3}(c) shows the unstable single-dimple equilibrium on the perfect odd snaking branch at the same end-shortening $\Delta_{\mathrm{SNI}}\approx0.558\Delta_c$. The close similarity between the equilibria in Figs.~\ref{fig:S3}(b) and~\ref{fig:S3}(c) suggests that the defect pins a single-dimple equilibrium already present in the perfect shell. From the continuous family of circumferentially translated states, the defect selects the one centered on it. This pinning preserves reflection symmetry and provides a direct connection between the prebuckling base branch and the odd snaking branch.

\begin{figure*}[ht!]
    \includegraphics[width=0.98\linewidth]{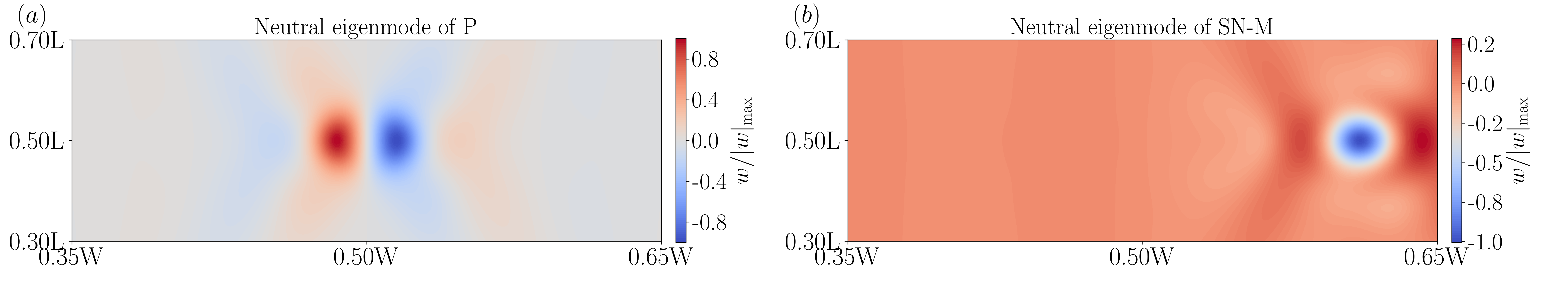}
    \caption{Neutral eigenmodes associated with (a) P, which connects the reflection-symmetric odd snaking branch and mixed branches, and (b) SN-M on the right-displaced mixed branch. The imperfect shell has $\delta=-0.50t$ and $l=0.55l_c$.}
    \label{fig:S4}
\end{figure*}

Figure~\ref{fig:S4}(a) shows the neutral eigenmode associated with the P bifurcation for the same imperfect shell in Fig.~1(b) of the main text. This mode displaces the central dimple circumferentially relative to the defect. It gives rise to two mixed branches whose equilibrium states are related by reflection about the defect center.
Figure~\ref{fig:S4}(b) shows the neutral eigenmode of SN-M on the mixed branch where the central dimple is displaced to the right of the defect. At the upper snaking limit $\Delta_+$, this mode nucleates a single dimple at the free front of the localized pattern.

\section{\centerline{4. Force--end-shortening bifurcation diagrams}}

\begin{figure*}[ht!]
    \includegraphics[width=\linewidth]{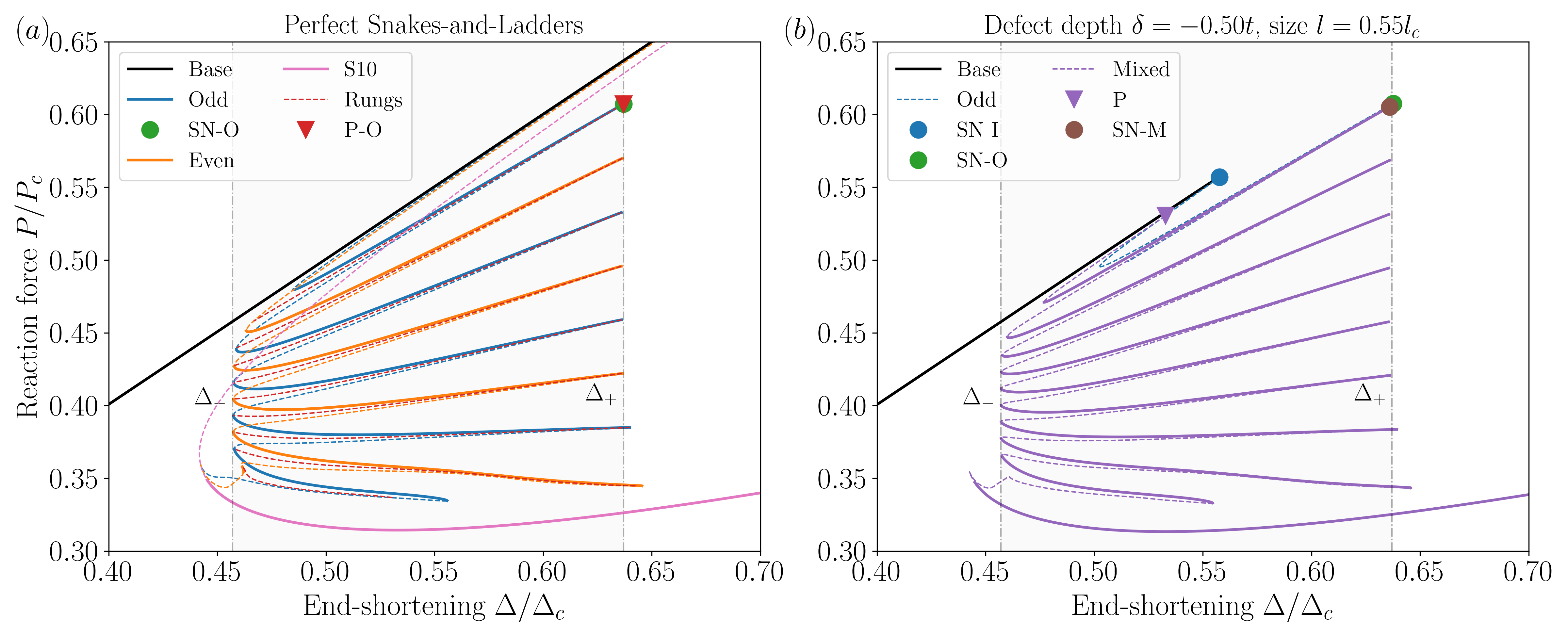}
    \caption{Bifurcation diagrams of the normalized reaction force, $P/P_c$, versus normalized end-shortening, $\Delta/\Delta_c$.
    (a) Perfect shell: base state, odd and even snaking branches, spatially periodic ten-dimple branch (S10), and rung branches.
    (b) Imperfect shell with $\delta=-0.50t$ and $l=0.55l_c$: base state, odd snaking branch, and mixed snaking branches. Branch and marker colors, and line styles follow those in Fig.~1 of the main text.}
    \label{fig:S5}
\end{figure*}

Figure~\ref{fig:S5} presents the bifurcation structures from Fig.~1 of the main text in terms of the normalized reaction force $P/P_c$ and normalized end-shortening $\Delta/\Delta_c$.
For the representative imperfect shell, the first loss of stability of the prebuckling state at SN~I is accompanied by an abrupt drop in reaction force. Stable localized equilibria on the mixed branches coexist with the stable prebuckling state over a finite interval of end-shortening and remain stable beyond SN~I. The static bifurcation diagram alone, however, does not determine which postbuckling state is dynamically selected following loss of stability. We therefore define the buckling threshold throughout the main text as the first loss of stability of the prebuckling state encountered under increasing end-shortening.

\bibliography{reference}